\documentclass[aps,prb,amsmath,amssymb,amstext,citeautoscript,punctuation,nofootinbib,superscriptaddress,twocolumn]{revtex4-2}
\usepackage{amsthm,mathrsfs,amsfonts,dsfont,hyperref,epsfig,braket,bm,enumerate,color,graphicx,dcolumn,charter,gensymb,comment,physics,cleveref,mathtools,natbib}
\usepackage[normalem]{ulem}
\usepackage[dvipsnames]{xcolor}
\usepackage [english]{babel}
\newcommand{\ecs}{EuCd$_2$Sb$_2$}
\begin{document}
\title{Magnetic structure of \ecs{} single-crystal thin-film} \date{\today}
\author{Eliot Heinrich}
\affiliation{Department of Physics, Boston College, 140 Commonwealth Avenue, Chestnut Hill, Massachusetts 02467, USA}
\author{Ayano Nakamura}
\affiliation{Department of Physics, Tokyo Institute of Technology, Tokyo 152-8551, Japan}
\author{Shinichi Nishihaya}
\affiliation{Department of Physics, Tokyo Institute of Technology, Tokyo 152-8551, Japan}
\author{Eugen Weschke}
\affiliation{Helmholtz-Zentrum Berlin für Materialien und Energie, Wilhelm-Conrad-Röntgen-Campus BESSY II, Albert-Einstein-Strasse 15, 12489 Berlin, Germany}
\author{Henrik Rønnow}
\affiliation{Institute of Physics, Ecole Polytechnique Fédérale de Lausanne (EPFL), CH-1015 Lausanne, Switzerland}
\author{Masaki Uchida}
\affiliation{Department of Physics, Tokyo Institute of Technology, Tokyo 152-8551, Japan}
\author{Benedetta Flebus}
\affiliation{Department of Physics, Boston College, 140 Commonwealth Avenue, Chestnut Hill, Massachusetts 02467, USA}
\author{Jian-Rui Soh}
\affiliation{A*STAR Quantum Innovation Centre (Q.InC), Institute of Materials Research and Engineering (IMRE), Agency for Science Technology and Research (A*STAR), 2 Fusionopolis Way, Singapore 138634}
\affiliation{Institute of Physics, Ecole Polytechnique Fédérale de Lausanne (EPFL), CH-1015 Lausanne, Switzerland}
\begin{abstract}
We investigate the magnetic order in single crystalline \ecs{} thin films using a combined theoretical and experimental approach. Resonant elastic x-ray scattering experiments reveal a sharp magnetic peak at $\bold{q}$$=$$(0,0,\frac{1}{2})$ below $T_\mathrm{N}=7.2$\,K, indicative of interlayer antiferromagnetic ordering. Additionally, we observe a weak diffuse magnetic signal centered at $\bold{q}$$=$$(0,0,1)$ that persists above $T_\mathrm{N}$, up to $T_\mathrm{C}\sim$11\,K. Our Monte-Carlo simulations of a classical spin model approximation of the Eu magnetic sublattice demonstrate that the diffuse signal can arise from ferromagnetic coupling in the top few layers due to surface oxidation. On the other hand, the bulk of the sample exhibits antiferromagnetic coupling between layers. Finally, our fit of the model parameters to the magnetic ordering temperatures, shed light on the exchange couplings that are key in stabilizing the observed composite magnetic order. 
\end{abstract}
\maketitle
\section{Introduction}
In 1929, Hermann Weyl theorized the existence of massless electrons, i.e., electrically charged particles predicted to travel at the speed of light~\cite{weyl_equation}. While these particles have not yet been observed in high-energy physics experiments, certain crystalline solids that fulfill specific symmetry requirements can realize massless fermionic quasiparticles and, thereby, Weyl's prediction. These materials are known as Weyl semimetals (WSMs), whose massless charge-carrying fermions can enable exceptionally high electron mobility, making these systems highly desirable for electronic devices~\cite{bernevig_progress_2022,tokura_magnetic_2019,wang_intrinsic_2023}.

In 2015, the existence of WSMs was experimentally confirmed, with several crystalline materials shown to host massless charge carriers characterized by linear crossings in the electronic spectrum, as predicted by Weyl~\cite{yang_weyl_2015,Su_Science_2015,lv_observation_2015}. These crossings, referred to as Weyl nodes, are stabilized by topological invariants of the electronic bands rather than, e.g., the symmetries that protect Dirac fermions in graphene~\cite{armitage_weyl_2018,RevModPhys.93.025002}. The topological nature of the Weyl nodes renders them exceptionally stable. Their stability and high electron mobility make them particularly attractive for device applications, though precise control over their properties remains challenging.

\begin{figure}[b!]
    \centering
\includegraphics[width=0.49\textwidth]{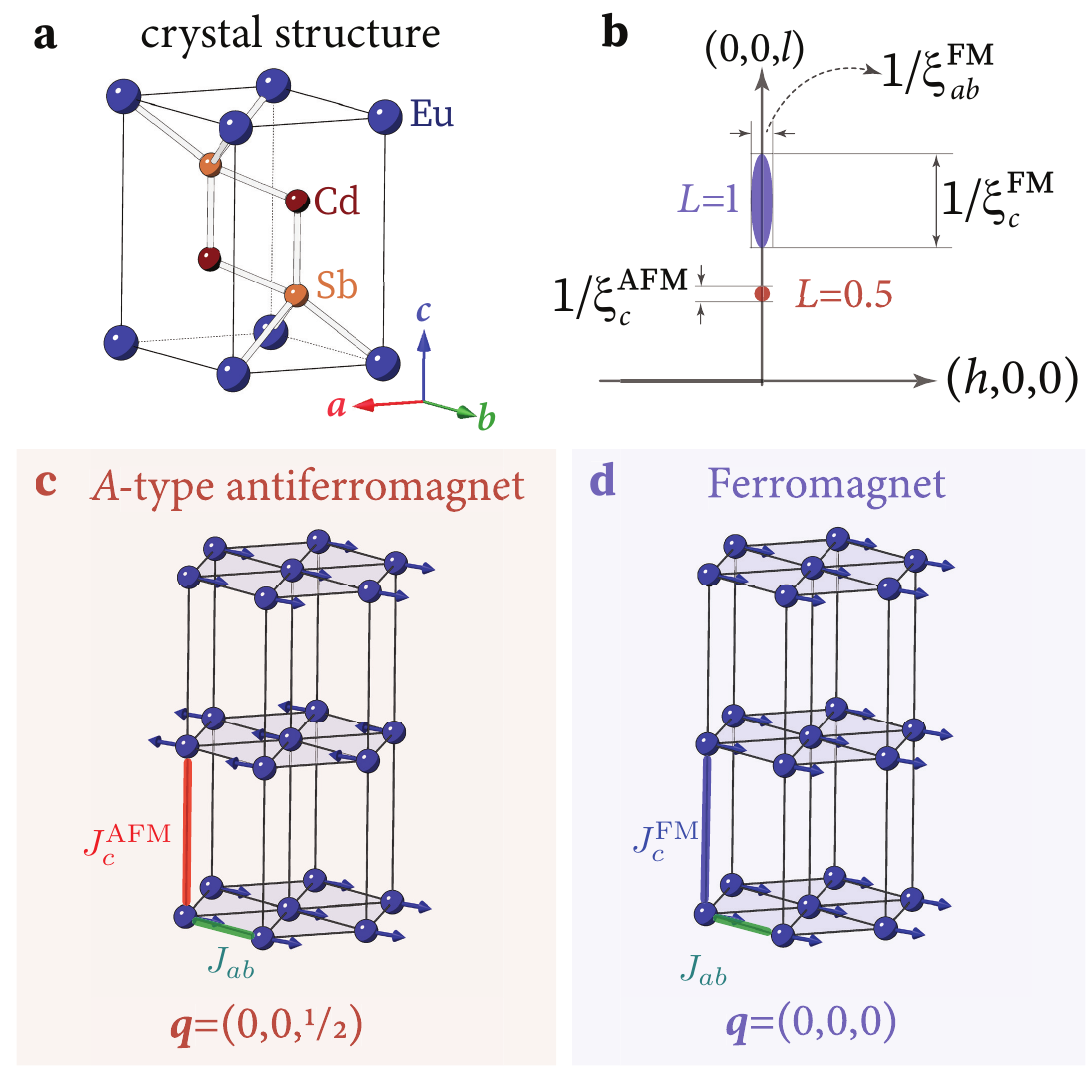}
    \caption{\textbf{a} The trigonal crystal structure of \ecs{}. \textbf{b} ($h$,0,$l$) reciprocal space map of \ecs{} around the magnetic diffuse (0,0,1) and magnetic sharp (0,0,$\frac{1}{2}$) peak. \textbf{c} $A$-type antiferromagnetic and \textbf{d} ferromagnetic order. \label{Crystal_Structure}}
\end{figure}

Recent work demonstrated that the presence or absence of Weyl nodes in a new class of materials called magnetic WSMs could be controlled by the symmetry of the magnetic configuration~\cite{Soh_Ideal_Weyl}.  This breakthrough shows that the mass of charge carriers in these magnetic WSMs can be tuned by modifying the topology of the electronic bands through external magnetic field modulation.

One such example is EuCd$_2$Sb$_2$, which features an intimate coupling between the magnetism of europium atoms and the topological properties of the bands near the Fermi energy. The trigonal unit cell of \ecs{}, shown in Fig.~\ref{Crystal_Structure}, possesses $P\bar{3}m1$ space group symmetry, which include a three-fold rotation symmetry about the $c$ axis and the global inversion symmetry. With europium (cadmium, antimony) residing on the $1a$ ($2d$) Wyckoff site, EuCd$_2$Sb$_2$ forms a layered structure, comprising alternating magnetic Eu and charge transport Cd$_2$Sb$_2$ layers. 

Crucially, their stacking leads to a strong interplay between charge carriers with non-trivial topology and magnetic order. However, despite its promising features, the full technological potential of \ecs{} remains largely untapped as it is currently available only in bulk crystal form. To leverage its properties for device applications, there is a pressing need to downscale EuCd$_2$Sb$_2$. Addressing this demand, single crystal thin films of EuCd$_2$Sb$_2$ have recently been successfully fabricated~\cite{masaki,PhysRevB.109.L121108}. However, several of their properties, including the ground-state magnetic order, remain inadequately explored.

In this work, we study the magnetic order of \ecs{} theoretically and experimentally via Monte Carlo (MC) simulations and resonant elastic x-ray scattering (REXS), respectively. We find that thin film \ecs{} displays $A$-type antiferromagnetic ordering, characterized by intralayer ferromagnetic coupling and interlayer antiferromagnetic interactions. Furthermore, we observe a broad peak in the x-ray scattering intensity around the $\bold{q} = (0,0,1)$ structural reflection. Our numerical simulations demonstrate that this broad peak is well-described by assuming that the interlayer coupling for the top few layers to be ferromagnetic, while the remaining layers exhibit antiferromagnetic coupling. We attribute the coexistence of ferromagnetic and antiferromagnetic interlayer order in \ecs{}  to surface oxidation.

\section{Methods}
Single-crystal thin film \ecs{} of high crystalline quality was grown on a cadmium telluride\,(111) substrate by means of molecular beam epitaxy (MBE) as outlined in detail in Ref.~\cite{masaki}. The MBE growth resulted in a layered \ecs{} sample, with the crystal $c$ axis oriented along the substrate surface normal. Correspondingly, the crystal $a$ and $b$ axes lie within the plane of the CdTe substrate. The thickness of the sample correspond to approximately 59 atomic layers along the $c$ axis. 

To investigate the magnetic properties of the thin film \ecs{}, temperature-dependent magnetic susceptibility measurements was performed with a Physical Properties Measurements System (Quantum Design) in various fixed magnetic field strengths, down to $T$\,=\,2\,K. Here, the external magnetic field was applied along the crystal $b$ axis. Since the CdTe substrate is diamagnetic, its contribution to the measured magnetic susceptibility manifests as a temperature-independent background signal, which can be subtracted from the raw data so as to isolate the signal arising from \ecs{}.

To determine the Eu magnetic order and how it develops with temperature, REXS measurements were performed on thin film \ecs{} on the UE46-PGM01 beamline~\cite{UE46_PGM1} at the BESSY II synchrotron facility. The incident soft x-ray photon energy was tuned to the europium $M_5$ absorbtion edge ($E_\mathrm{res.}$=1128.8\,eV), so as to benefit from the resonant enhancement of the scattered x-ray intensity arising from the ordering of the Eu$^{2+}$ magnetic moments~\cite{hill_REXS}. The main drawback associated with operating at $E_\mathrm{res.}$ is that the only accessible structural peak in the whole reciprocal space of \ecs{} --where the incident or scattered beam is not otherwise blocked by the sample itself-- is the $\textbf{\textit{Q}}$=($0,0,1$) reflection. 

Notwithstanding, the information regarding the Eu magnetic order can still be acquired by studying the scattered REXS signal in the vicinity of the ($0,0,1$) peak. Fortuitously, the scattering angle ($2\theta$) of the ($0,0,1$) peak is close to 90$^\circ$, which allows for the suppression of the charge scattering signal when the incident x-rays with $\pi$ linear polarization is used, as was the case for our REXS experiment. In particular, the \ecs{} thin film sample was mounted with the crystal $b$ axis perpendicular to the horizontal scattering plane of the diffractometer, in order to access the ($h,0,l$) reciprocal space area in the vicinity of ($0,0,1$) peak, as shown in Fig.~\ref{Crystal_Structure}\textbf{b}.

We performed MC simulations to model the temperature evolution of the Eu magnetic order obtained from the REXS measurements. The magnetic Hamiltonian of the layered trigonal lattice of EuCd$_2$Sb$_2$ can be written as~\cite{heinrich_model_2022}

\begin{align}
    \mathcal{H} = &-J_{ab} \sum\limits_{\langle i, j \rangle_{ab}} \mathbf{S}_{i} \cdot \mathbf{S}_{j} + \sum\limits_{\langle i, j \rangle_c} J_c^{ij} \mathbf{S}_i \cdot \mathbf{S}_{j} \nonumber \\
    &+ K_2 \sum\limits_{i} \cos^2(\theta_i) + K_6 \sum\limits_{i} \sin^6(\theta_{i}) \cos(6\phi_{i}),
    \label{Hamilt}
\end{align}
where $\mathbf{S}_{i} = \left(\cos(\phi_{i})\sin(\theta_{i}), \sin(\phi_{i})\sin(\theta_{i}), \cos(\theta_{i}) \right)$ describe the spin magnetic moment of an Eu$^{2+}$ ion at the lattice site $\textbf{r}_{i}$, whose polar ($\theta_{i}$) and azimuthal ($\phi_{i}$) angles are defined in a Cartesian coordinate system with the $\mathbf{z}$ axis aligned along the crystal $c$ axis. The Eu$^{2+}$ ions have a spin moment of $S = 7/2$, but here we set $S = |\mathbf{S}_{i}| = 1$ and absorb the factor $ 7/2$ into the definitions of the coupling coefficients $J_{ab}$, $J_c^{ij}$, $K_2$, and $K_6$. The sum $\langle \cdot \rangle_{ab}$ indicates a sum over nearest neighbor spins in the $ab$ plane, namely pairs of intralayer nearest neighbors. Likewise, the sum $\langle \cdot \rangle_c$ indicates a sum over nearest neighbor spins along the $c$ axis, between pairs of interlayer nearest neighbors.


The parameters $J_{ab}$ and $J_c$ correspond to the in-plane and out-of-plane Heisenberg-like exchange coupling, respectively [Fig.~\ref{Crystal_Structure}\textbf{c}, \textbf{d}]. Note that the interlayer coupling $J_c^{ij}$ has a dependence on the sites $i$ and $j$, and is thus allowed to vary over the thickness of the sample. In particular, we choose
\begin{align*}
	J_c^{ij} = 
	\begin{cases}
		-J_c^\mathrm{FM} & i,j \in \text{top }\ell \text{ layers} \\
		J_c^\mathrm{AFM} & \text{else}
	\end{cases}.
\end{align*}

We also include an easy-plane anisotropy with strength $K_2 > 0$ and a sixfold magnetocrystalline anisotropy with strength $K_6 > 0$, as observed in similar materials \cite{heinrich_model_2022, wang_eucd2p2_2021}. The REXS intensity along the $\textbf{Q}$=$(0,0,l)$ direction arising from a given magnetic configuration 
$\{\mathbf{S}_{i}\}$  is computed by taking the sum,
\begin{align}
	I(l) = \frac{1}{N} \left\vert \sum\limits_{i} e^{-2\pi \mathrm{i}l }  \mathbf{S}_{i} \right\vert^2.
\end{align}


To perform the MC simulation, we employ Wolff cluster updates \cite{wolff_cluster_1989} with simulated annealing \cite{kirkpatrick_annealing_1983} to reach statistical equilibrium. To account for the onsite anisotropic fields, we employed the `ghost spin' technique presented in Ref. \cite{kent_externalfield_2018}. We generated random spin reflections for the Wolff cluster updates by sampling a Gaussian random $\boldsymbol \Gamma$, where the vector elements $\Gamma_i \sim N(0, 1)$ for $i \in \{x, y, z\}$ and $N(\mu, \sigma)$ is a normal distribution of mean $\mu$ and standard deviation $\sigma$. This vector $\boldsymbol \Gamma$ is used to generate the reflection matrix $R = \mathbb{I}_{3 \times 3} - 2\boldsymbol \Gamma \boldsymbol \Gamma^T / |\boldsymbol \Gamma |^2$, where $\mathbb{I}_{3 \times 3}$ is a $3 \times 3$ identity matrix.

\begin{figure}[b!]
    \centering
    \includegraphics[width=0.45\textwidth]{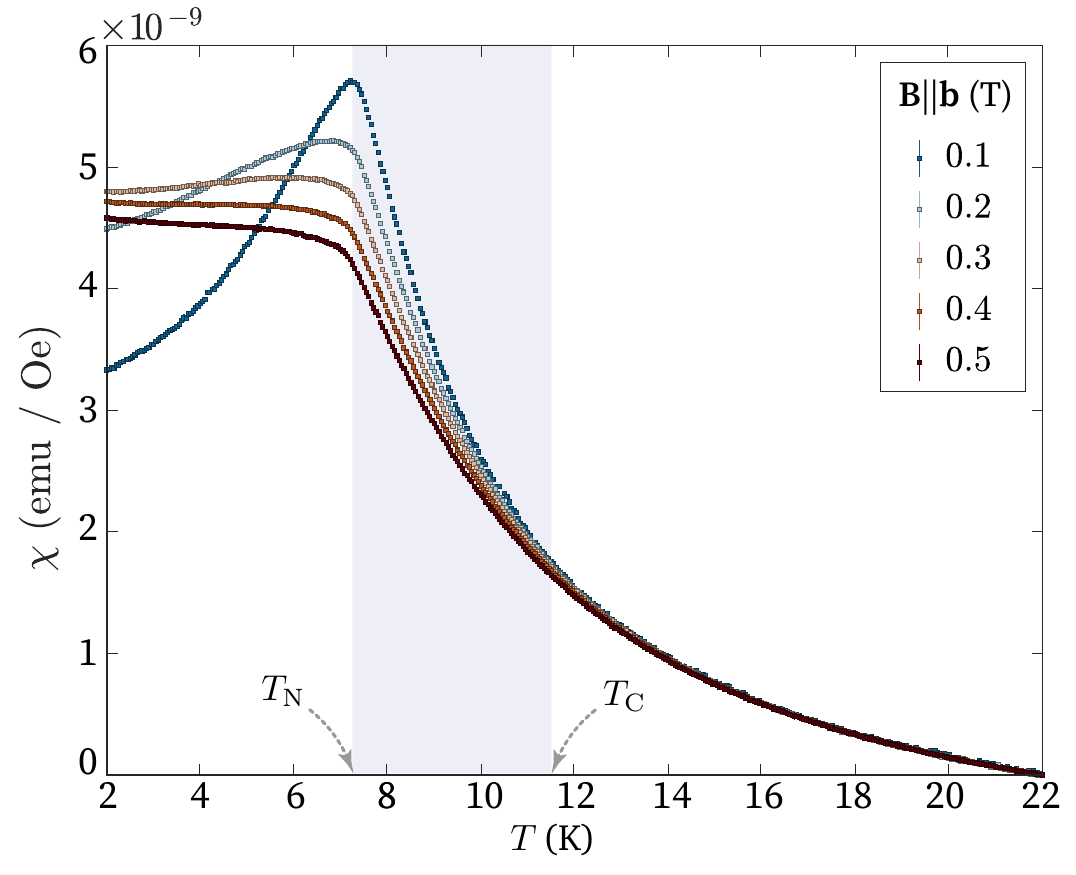}
    \caption{The in-plane magnetic susceptibility of \ecs{} as a function of temperature, in the presence of various external magnetic field along the crystal $b$ axis, display a clear anomaly at $T_\mathrm{N}\sim7.2$\,K. The temperature-independent background contribution from the CdTe substrate \cite{CdTe_sus} was subtracted by normalizing the magnetic susceptibility at $T$$=$$22$\,K to zero. Indeed, for temperatures above 14\,K, we find that the susceptibility curves coincide with each other. \label{chi-T}}
\end{figure}

At a given temperature $T$, we estimated the intensity $\langle I(l) \rangle = I(l, T)$ by performing the following protocol:
\begin{enumerate}
    \item Randomly initialize each spin on the unit sphere.
    \item Perform $n_\text{max} = 2000$ cluster updates, with the temperature at the $n$th step updated according to the annealing schedule 
    \begin{align}
        T_n = T_i + \frac{T_f - T_i}{2} \left(1 - \cos(\pi n/n_\text{max})\right)\,,
    \end{align}
    where $T_i = 2.5 J/k_\mathrm{B}$ and $T_f = T$ is the temperature of the ensemble we wish to sample. 
    \item Perform $N_\text{samples} = 2000$ cluster updates at fixed temperature $T$, sampling the intensity after every $20$th update. Record these samples as $I_j(l,T)$.
    \item Report the ensemble averages 
    \begin{align}
    (l,T) \approx \frac{1}{N_\text{samples}} \sum\limits_j I_j(l,T).
    \label{MC}
    \end{align}
\end{enumerate}

\begin{figure*}[t!]
\centering
\includegraphics[width=0.99\textwidth]{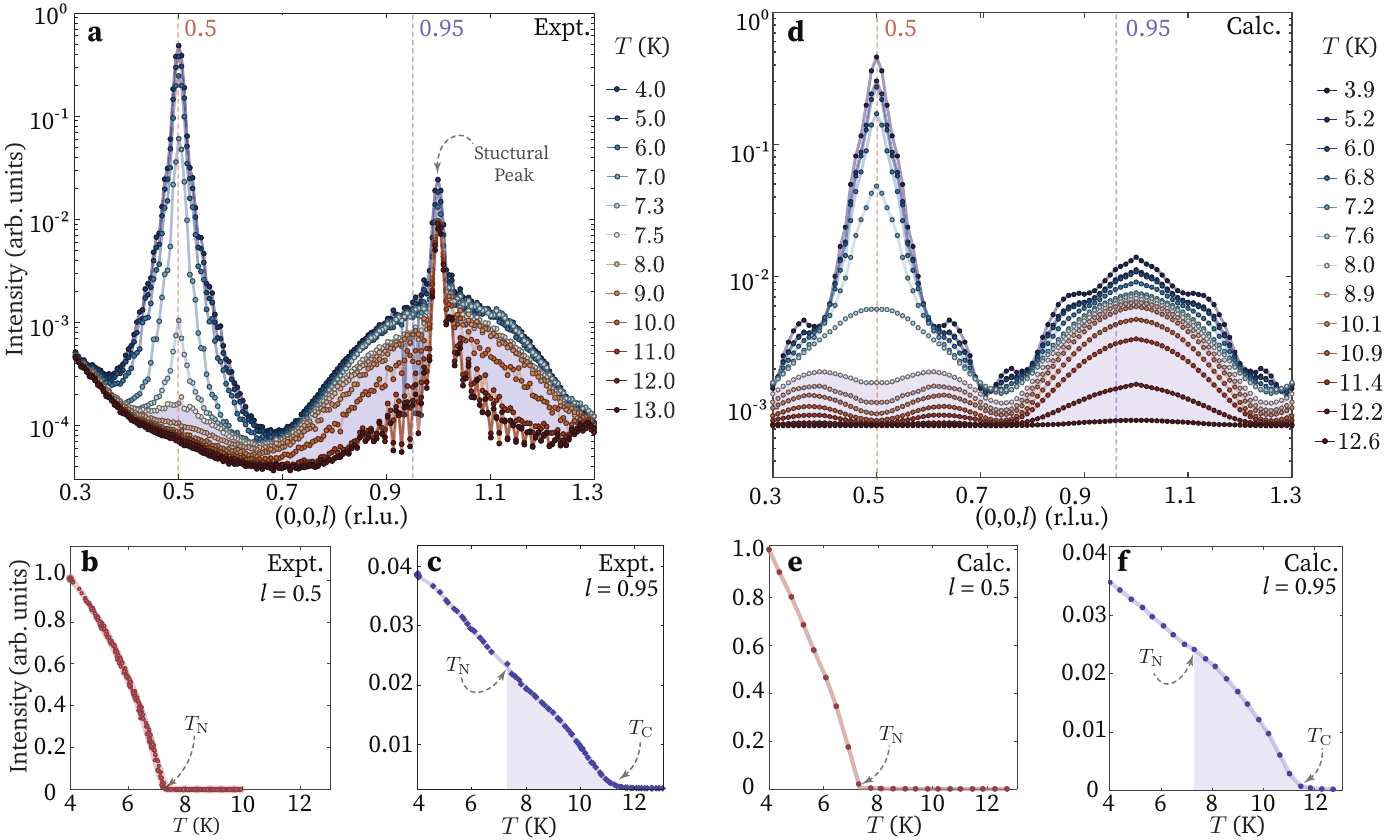}
\caption{ Comparison between the (\textbf{a}--\textbf{c}) experimental and (\textbf{d}--\textbf{f}) calculated REXS intensity as a function of temperature. \textbf{a},\textbf{d} The intensity along (0,0,$l$) at various temperatures. \textbf{b},\textbf{e} The antiferromagnetic peak intensity at $l$=0.5. \textbf{c}, \textbf{f} The temperature dependence at $l$=0.95 to probe the diffuse ferromagnetic signal. ~\label{lscan-T}}
\end{figure*}

\section{Results}
Figure~\ref{chi-T} shows the temperature-dependent magnetic susceptibility curves obtained at various external field strengths.  At low-field strengths ($\mu_0H$=0.1\,T), the magnetic susceptibility curve displays a sharp anomaly at $T_\mathrm{N}$=$7.2(1)$\,K, which is indicative of the onset of long-range AFM order of the europium magnetic moments. The value for $T_\mathrm{N}$ is slightly lower compared to those found for bulk crystalline \ecs{} of $T_\mathrm{N}$=$7.4$\,K~\cite{soh_magnetic_2018}. This discrepancy  can be attributed to variations in the strength of magnetic exchange coupling, stemming from differences in the size of the cell parameters between single crystals and thin films.

 To ascertain if thin film \ecs{} possess the $A$-type AFM order of the europium moments observed in the bulk crystal~\cite{soh_magnetic_2018}, we proceed to analyze the low-temperature REXS data [Fig.~\ref{lscan-T}\textbf{a}--\textbf{c}]. The presence of an $A$-type AFM configuration, i.e., ferromagnetic europium layers  stacked alternately along the crystal $c$ axis [Fig.~\ref{Crystal_Structure}\textbf{c}], is confirmed by the strong Bragg reflection observed at $T$$\sim$$4$K for $\textbf{\textit{Q}}$=$\left(0,0,\frac{1}{2}\right)$, which is forbidden by the $P\bar{3}m1$ space group of \ecs{}. Such a peak arises from the doubling of the unit-cell along the crystal $c$ axis, which is consistent with an $A$-type AFM order [Fig.~\ref{Crystal_Structure}\textbf{a}]. Further confirmation is found in the temperature dependence of the $\left(0,0,\frac{1}{2}\right)$ peak [Fig.~\ref{chi-T}\textbf{c}], which disappears on warming above $T_\mathrm{N}$=7.2\,K concomitant with the anomaly observed in the magnetic susceptibility curves.

We now turn our focus to investigating whether thin film \ecs{} can also exhibit other types of europium magnetic configurations coexisting with the AFM order. To this end, we measured the REXS intensity along the $\textit{\textbf{Q}}$=$(0,0,l)$ direction in reciprocal space, which is sensitive to other types of stacking arrangements of ferromagnetic europium layers along the crystal $c$ axis. For instance, one possibility is a ferroic stacking of the ferromagnetic Eu layers, as shown in Fig.~\ref{Crystal_Structure}\textbf{d}. Such a ferromagnetic (FM) order will produce magnetic scattering intensity centered at $\textbf{\textit{Q}}$=$(0,0,1)$ in addition to the AFM peak at $(0,0,\frac{1}{2})$.

Figure~\ref{lscan-T}\textbf{a} shows the scattered x-ray intensity along $\textit{\textbf{Q}}$=$(0,0,l)$ measured at various temperatures. In addition to the  AFM peak at $l$=0.5 at temperatures  below $T_\mathrm{N}$, we observe the sharp charge peak at $l$=1 arising from the (0,0,1) reflection of the crystal structure. Interestingly, between $l$=0.7 and 
 $l$=1.3, we also detect a broad domed-shaped diffuse signal centred around the sharp (0,0,1) structural peak.
 
 To ascertain the temperature at which this diffuse scattering starts to develop, we track the temperature dependence of the REXS signal at $l$=0.95 [Fig.~\ref{lscan-T}\textbf{c}]. Studying the signal at $l$=0.95 allows us to fulfil Brewster's condition, where the scattering angle ($2\theta$) is 90$^\circ$, whilst also avoiding the strong structural peak at $l$=1. As such, the charge scattering is strongly suppressed since incident x-rays with $\pi$ linear polarization were used. Therefore, the resonant x-ray signal detected at $l$=0.95 is predominately due to the magnetic ordering, which rotates the incident x-ray polarization into the $\pi$$\to$$\sigma^\prime$ scattering channel. Strikingly, we find that the diffuse magnetic signal already starts to develop at $T_\mathrm{C}$$\sim$11\,K, a temperature well above $T_\mathrm{N}$, as shown in Fig.~\ref{lscan-T}\textbf{c}. 

The shaded regions in Figs.~\ref{lscan-T}\textbf{a}, \textbf{c} denote the changes in the REXS intensity between $T_\mathrm{N}$ and $T_\mathrm{C}$, which clearly highlight that the diffuse magnetic signal grows even in the absence of AFM order. Since the diffuse signal in the shaded region of Fig.~\ref{lscan-T}\textbf{a} is centered at $l$=1, it can be attributed to the spontaneous ferromagnetic stacking of europium layers along the crystal $c$ axis, as shown in Fig.~\ref{Crystal_Structure}\textbf{d}. The number of layers involved in this ferromagnetic (FM) order should be fairly limited, given that the peak width along the $c$ axis is very broad [Fig.~\ref{Crystal_Structure}\textbf{b}]. Indeed, our fit finds that the correlation length of this FM order along the crystal $c$ axis ($\xi_c^\mathrm{FM}$) corresponds an ordering of approximately three basal layers. In contrast, the correlation length within the $ab$ plane is much larger, i.e., of the order of $\xi_{ab}^\mathrm{FM}$$\sim$1230\,\AA{} at $T$=4\,K.

 A natural question that arises is whether the two types of magnetic order are coupled, that is, for instance, if the onset of the AFM order at $T_\mathrm{N}$ enhances or suppresses the three-layered  FM order. Notably, we find that the broad peak in the $(0,0,l)$ scan associated with the FM order retains its dome-shaped signal even below $T_\mathrm{N}$, as shown in Fig.~\ref{lscan-T}\textbf{a}. This behavior demonstrates that the FM order not only persists below $T_\mathrm{N}$, but also retains its three-layered thickness, suggesting that the two types of magnetic orders are largely decoupled. This conclusion is corroborated by the temperature dependence at $l$=0.95 in Fig.~\ref{lscan-T}\textbf{c}, which does not exhibit a significant jump at $T_\mathrm{N}$ in the $l$=0.5 curve [Fig.~\ref{lscan-T}\textbf{b}]. 

Yet, our experimental findings prompt further questions regarding the mechanism responsible for the emergence of a ferromagnetic interlayer exchange coupling, and the factors that confine this mechanism to only a few top layers.
 \begin{figure}[t!]
\centering
\includegraphics[width=0.49\textwidth]{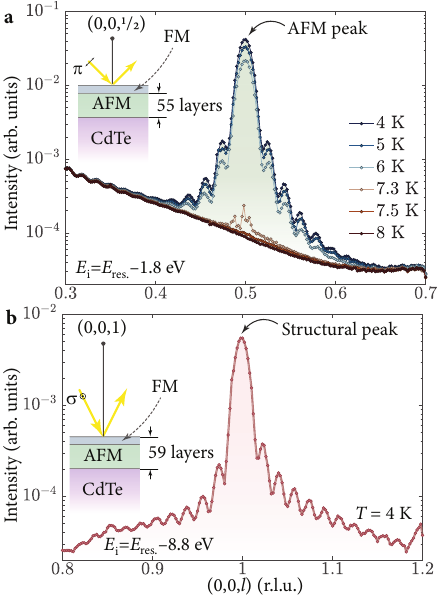}
\caption{ Laue oscillations of the antiferromagnetic (0,0,$\frac{1}{2}$) and structural (0,0,1) peak. \textbf{a} REXS intensity of the (0,0,$\frac{1}{2}$) peak measured slightly off-resonance ($E_\mathrm{i}$=$E_\mathrm{res.}$-1.8\,eV) to determine the thickness of the AFM layer. \textbf{b}, Non-resonant x-ray scattering intensity ($E_\mathrm{i}$=$E_\mathrm{res.}$-8.8\,eV) of the (0,0,1) structural peak to ascertain the total thickness of EuCd$_2$Sb$_2$.~\label{Laue_Osc}}
\end{figure}
A plausible explanation is that the EuCd$_2$Sb$_2$ sample studied in this work suffers from surface oxidation, albeit  limited to few top atomic layers. This phenomenon is common in europium-based inter-metallics, whose surface oxidation leads to the formation of a protective layer of Eu$_2$O$_3$ preventing the underlying sample from  further damage\cite{Eu2O3,Eu2O3_2}. Due to oxidation, some of the magnetic Eu$^{2+}$ becomes a non-magnetic Eu$^{3+}$ ion. As a result, the antiferromagnetic exchange path between neighboring ions along $c$ can change sign, becoming ferromagnetic~\cite{PhysRevB.101.140402}. This mechanism also limits ferromagnetically coupled layers to the top few layers, while preserving the AFM order of the underlying layers. 

In order to verify this hypothesis, it is necessary to determine the thickness of the AFM layer. This can be achieved by measuring the Laue oscillations associated with the (0,0,$\frac{1}{2}$) peak~\cite{Laue_osc_1}. However, operating at the peak resonant energy $E_\mathrm{res.}$ leads to strong self absorption of the scattered signal that tends to wash out these subtle oscillations. Hence, we performed the REXS measurements at a slightly off-resonance condition ($E_\mathrm{i}$=$E_\mathrm{res.}$$-$1.8\,eV). This incident photon energy sits at the tail of the Eu$^{2+}$ resonance, ensuring that the probe remains sensitive to the signal  originating from the AFM order.

Figure~\ref{Laue_Osc}\textbf{a} shows the Laue oscillation of the scattered signal arising from the AFM coupled layers. Our fits indicate that the signal comes from $\sim$55 AFM coupled europium layers, which, when summed with the three  FM coupled top layers, yield a total thickness of  58 layers. This estimate is in good agreement with the 59-layer thickness of EuCd$_2$Sb$_2$ extracted from the Laue oscillations of the (0,0,1) structural peak measured in the non-resonant regime of $E_\mathrm{i}$=$E_\mathrm{res.}$-8.8\,eV [Fig.~\ref{Laue_Osc}\textbf{b}].

To shed further light on the magnetic properties of \ecs{} single-crystal thin-film, we performed Monte Carlo simulations on a system consisting of 32 layers, with each layer containing $64 \times 64$ ions.  The top 3 layers are ferromagnetically (FM) coupled, while the other layers are antiferromagnetically (AFM) coupled, with exchange strengths $J_{c}^{\mathrm{FM}}$ and $J_{c}^{\mathrm{AFM}}$ respectively. 

The Heisenberg-like exchange coupling ($J_{ab}$, $J_{c}^{\mathrm{FM}}$. $J_{c}^{\mathrm{AFM}}$) and magnetocrystalline anisotropy ($K_2$, $K_6$) parameters [Eq.~\ref{Hamilt}] are chosen to provide the best qualitative and numerical fit to the REXS scattering data.

Figure~\ref{lscan-T}\textbf{d}--\textbf{f} shows the calculated REXS intensity for $J_{ab}$=0.54\,meV, $J_c^\mathrm{AFM}$=0.011\,meV, $J_c^\mathrm{FM}$=0.54\,meV, $K_2$=0.14\,meV and $K_6$=0.11\,meV. Despite the finite size of the magnetic supercell, the temperature dependence calculated in the numerical simulations [Figs.~\ref{lscan-T}\textbf{d}--\textbf{f}] is in good agreement with the  corresponding experimental data shown in Figs.~\ref{lscan-T}\textbf{a}--\textbf{c}. For instance, the calculated ordering temperatures of $T_\mathrm{N}$=7.2\,K and $T_\mathrm{C}$=11.4\,K [Figs.~\ref{lscan-T}\textbf{e}, \textbf{f}], are in good accordance with the experimentally observed values $T_\mathrm{N}$=7.2\,K and $T_\mathrm{C}$=11\,K [Figs.~\ref{lscan-T}\textbf{b}, \textbf{c}], respectively. Furthermore, the calculated dome-shaped diffuse signal centered around (0,0,1) in Figs.~\ref{lscan-T}\textbf{a} agrees with that of the experimentally obtained plot [Fig.~\ref{lscan-T}\textbf{d}].


\section{Conclusion}
In this work, we have shown through theoretical and experimental means that thin film \ecs{} exhibits $A$-type antiferromagnetic ordering, coexisting with ferromagnetic ordering in the few top  layers. By setting the interlayer coupling in the surface layers to be ferromagnetic, we find close qualitative and quantitative agreement between our REXS and MC data. We find that the ordering temperatures of the ferromagnetic ordering and $A$-type antiferromagnetic ordering differ significantly, indicating that these coexisting magnetic orders are uncoupled. This coexisting magnetic ordering may be explained by surface oxidation of the sample, producing a film of Eu$_2$O$_3$ and changing the strength and sign of the magnetic exchange near the surface of the sample. 


Given that interlayer and intralayer exchange interactions are both present, the system is essentially behaves as a 3d system from a magnetic point of view. In future investigations, it would be interesting to understand how the magnetism develops as the thickness of the sample approaches the monolayer limit, to explore phenomena such as possible Berezinskii-Kosterlitz-Thouless (BKT) transitions predicted in Refs.~\cite{heinrich_model_2022,PhysRevB.104.L020408}. To that end, it will be crucial to explore how surface oxidation can be limited so as to avoid the formation of coexisting magnetic orders. 

\begin{acknowledgments}
We are grateful for the helpful contributions of X.-P. Yang and also acknowledge the BESSY II facility of the Helmholtz-Zentrum Berlin für Materialien und Energie for the provision of the beamtime under proposal No. 222-11565-ST. B.F. acknowledges support from the National Science Foundation under Grant No. NSF DMR-2144086. J.-R.S. acknowledges support from the Singapore National Science Scholarship, Agency for Science Technology and Research and the European Research Council (HERO, Grant No. 810451). The MC calculations were done with computational support from the Andromeda computing cluster at Boston College.
\end{acknowledgments}
\bibliography{bib.bib}
\end{document}